# Is Bayes theorem applicable to all quantum states?

**H. Razmi** [(1)] and **J. Allahyari** [(2)]

(1) Department of Physics, University of Qom, 3716146611, Qom, I. R. Iran.

razmi@qom.ac.ir & razmiha@hotmail.com

(2) Islamic Azad University, Khomeinishahr Branch, 84175-119, Khomeinishahr, Isfahan, I. R. Iran.
jahanshahallahyari@gmail.com

## Abstract

Reconsidering the already known important question that whether all the axioms and theorems in classical theory of probability are applicable to probability functions in quantum theory, we want to show that the so-called Bayes theorem isn't applicable to nonfactorizable quantum entangled states.

## Introduction

Although the probability theory has been known in classical physics for a long time with well known concept and applications; but, in quantum mechanics (QM), in spite of using it as an important and key issue, its fundamental concept and interpretation is still controversial. Among a number of research works on the validity of using classical theory of probability (CTP) in quantum theory (e.g. [1-3]), one of the important considerations is the study of (in)consistency of CTP and QM. Although some researchers has already tried to show there is complete consistency between QM and CTP [4], here, we are going to show there isn't such a complete consistency for all quantum states.

After a short review of Bayes theorem, we pay to quantum entanglement in summary and then show there isn't any possibility of applying Bayes theorem to quantum nonfactorizable (entangled) states.

## A short review of Bayes theorem

It is common to consider some events as elementary events (e.g. event $A$). There are some well-known notations for composite events such as $\sim A$ (not $A$) which shows the nonoccurrence of $A$, $A \wedge B$ ($A$ and $B$), which denotes the occurrence of both $A$ and $A$, $A \vee B$ ($A$ or $B$), which means occurrence of at least one of them. The operators ($\sim, \wedge, \vee$) are negation, conjunction, and disjunction respectively. The conditional probability function $P(A|B)$ meaning as the probability of the occurrence of event $A$ conditional on the occurrence of event $B$ has a basic role in introducing axioms of CTP. Among several equivalent expressions for the axioms of CTP and their corresponding theorems [5], considering the book by R. T. Cox [6], Bayes theorem is stated as the following:

$$P(A \wedge B) = P(A) P(B|A) \qquad (1).$$

This means that the conjunction probability of two events $A$ and $B$ is equal to the product of the probability of the occurrence of one event on the probability of the occurrence of the other one conditional on the occurrence of the first one. For two independent (mutually exclusive or stochastically/statistically exclusive) events, $P(B|A) = P(B)$; thus, using Bayes theorem, it is found that

$$P(A \wedge B) = P(A) P(B) \qquad (2).$$

## Nonfactorizable states and quantum entanglement

In classical physics, when two or more particles are combined and/or dependent on each other via any way as different forms of interaction and correlation, they can be separated and separately described particularly by enough spatial separation of them; but, in QM, there are some entangled states of two or more particles for them there isn't possible to have independent description and/or separated state of any of the particles even with distantly infinite spatial separation [7]. To explain more about such entangled states, consider tensor product of two Hilbert spaces $H^A$ and $H^B$ which is itself another Hilbert space may be named $H^{AB}$:

$$H^{AB} = H^A \otimes H^B \qquad (3).$$

For two state vectors $|\varphi^A\rangle \in H^A$ and $|\chi^B\rangle \in H^B$, there is a state vector:

$$|\varphi^A, \chi^B\rangle = |\varphi^A\rangle \otimes |\chi^B\rangle = |\varphi^A\rangle|\chi^B\rangle \in H^{AB} \qquad (4).$$

This is a factorizable combined state vector. The corresponding quantum mechanical density operator (matrix) is also decomposable as:

$$\rho^{AB} = \rho^A \rho^B \qquad (5);$$

and, one can simply check the applicability of Bayes theorem to such states.

But, there are some other combined states in the product space $H^{AB}$ which cannot be factorized as in the above relations. Among many ones, one can consider the following (singlet state) entangled state:

$$|\varphi, \chi\rangle^{AB} = \frac{1}{\sqrt{2}} (|\varphi^A\rangle|\chi^B\rangle - |\varphi^B\rangle|\chi^A\rangle) \qquad (6).$$

There isn't any possibility of decomposing the above entangled state as what happened in (4) and (5) and thus:

$$\rho^{AB} \neq \rho^A \rho^B \qquad (7).$$

**Is Bayes theorem applicable to all quantum states?**

As mentioned in the introduction, there are some efforts to show complete consistency between the axioms of CTP and QM; but, if one considers quantum entangled (nonfactorizable) states, there isn't such a consistency. One can simply check this by investigating the above-mentioned references that deal with the probability relations as $P(M_R \wedge M_S) = tr(\rho M_R M_S)$ which are only applicable to factorizable states. If we want to study the applicability of Bayes theorem to all quantum states, we should consider nonfactorizable quantum entangled states too. This is because these states don't follow the simple factorizable relations used in proving the consistency of Bayes theorem with quantum probability functions. As an example, consider the well-known singlet state:

$$|\Psi_{RS}\rangle = \frac{1}{\sqrt{2}}(|+\rangle_R|-\rangle_S - |-\rangle_R|+\rangle_S) \qquad (8).$$

As stated above, there isn't any possibility of decomposing/factorizing the density state of such an above quantum state in terms of $R$ and $S$ states and thus no any possibility of factorizing the conjunction probability $P_{RS}$ as a simple product of probabilities $P_R$ and $P_S$:

$$P(R \wedge S) = P(R)P(S) \qquad ? \qquad (9).$$

In different versions of Bell's theorem (e.g. the Clauser-Horne Model [8]), it is well-known that the so-called locality condition (9) cannot be applied to the entangled state (8).

Since the above locality condition is written based on Bayes theorem relation [9-10]

$$P(R \wedge S|\rho_{RS}) = P(R|\rho_{RS})P(S|R \wedge \rho_{RS}) \qquad (13),$$

this means, one cannot apply Bayes theorem to quantum nonfactorizable/entangled (here singlet state) states.

**Conclusion**

In this paper, we have shown there isn't a complete consistency between CTP and QM. This has been proved by checking the applicability of Bayes theorem to all quantum states and considering the inconsistency of quantum entangled states with this theorem. We think the inconsistency of quantum nonfactorizable (entangled) states with the so-called locality condition in different models of Bell's theorem (e.g. Clauser-Horne model) originates from the inconsistency between the CTP and quantum probability. Meanwhile, the reasons behind already known claims on the full consistency between the CTP and quantum probability are all based on working only with factorizable states rather than considering all quantum states consisting of quantum nonfactorizable (entangled) ones.